\documentclass[aps,pre,twocolumn]{revtex4}
\usepackage{graphicx}
\usepackage{amsmath}
\usepackage{array}
\usepackage{mathptmx}
\usepackage{anyfontsize}
\usepackage{t1enc}
\def\la{\left\langle\rule{0pt}{3em}}
\def\ra{\right\rangle}
\usepackage[colorlinks,linkcolor={blue},citecolor={blue},urlcolor={blue}]{hyperref}
\begin{document}

\title {Single particle fluctuations dominate the long-time dynamic
  susceptibility in glass-forming liquids.}
\author{Rajib K Pandit$^{1}$\footnote{Present address: Exact Sciences,
    Redwood City, 94063 CA, USA }} 
\author{Elijah Flenner$^{2}$}

\author{Horacio E. Castillo$^{1}$}
\email[Corresponding author, ]{castillh@ohio.edu}
\affiliation{$^{1}$Department of Physics and Astronomy and Nanoscale
  and Quantum Phenomena Institute, Ohio University, Athens, Ohio 45701,
  USA\\ 
$^{2}$Department of Chemistry, Colorado State University, Fort Collins, Colorado 80523, USA}

\date{\today}

\begin{abstract}
  Liquids near the glass transition exhibit dynamical heterogeneity,
  i.e. correlated regions in the liquid relax at either a much faster
  rate or a much slower rate than the average. This collective
  phenomenon has been characterized by measurements of a dynamic
  susceptibility $\chi_4(t)$, which are sometimes interpreted in terms
  of the size of those relaxing regions and the intensity of the
  fluctuations.
  We show that the results of those measurements can be affected not
  only by the collective fluctuations in the relaxation rate, but also
  by density fluctuations in the initial state and by single-particle
  fluctuations. 
  We also show that at very long times the average overlap $C(t)$
  probing the similarity between an initial and a final state
  separated by a time interval $t$ decays as a power law $C(t) \sim
  t^{-d/2}$. This is much slower than the stretched exponential
  behavior $C(t) \sim {\rm e}^{-(t/\tau)^{\beta}}$ previously observed
  at times within one or two orders of magnitude of the
  $\alpha$-relaxation time $\tau_{\alpha}$. 
  We find that for times longer than $10-100 \tau_{\alpha}$, the
  dynamic susceptibility $\chi_4(t)$ is dominated by single particle
  fluctuations, and that $\chi_4(t) \approx C(t) \sim t^{-d/2}$.
  Finally, we introduce a method to extract the collective relaxation
  contribution to the dynamic susceptibility $\chi_4(t)$ by
  subtracting the effects of single-particle fluctuations and initial
  state density fluctuations. We apply this method to numerical
  simulations of two glass forming models: a binary hard sphere system
  and a Kob-Andersen Lennard-Jones system. This allows us to extend
  the analysis of numerical data to timescales much longer than
  previously possible, and opens the door for further future progress
  in the study of dynamic heterogeneities, including the determination
  of the exchange time.
\end{abstract}

 
\maketitle

\section{Introduction}
\label{sec:introduction}
Glass forming liquids are characterized by a dramatic slowdown of the
relaxation dynamics as the temperature is reduced or the density is
increased. A common way to probe relaxation is to measure the
similarity between states of the system at different times. To do this
an often used quantity is the average overlap $C(t)= \langle w(\Delta
\vec{r}) \rangle$. Here $w(\Delta \vec{r})$ is an individual particle
overlap function that goes from one to zero as the particle
displacement $\Delta \vec{r} (t)$ goes from being smaller to being
larger than a typical vibrational
amplitude~\cite{Lacevic2003,Berthier2011}. The main timescale
describing the slowdown of the dynamics is the $\alpha$-relaxation
time $\tau_{\alpha}$. This timescale characterizes the decay of an
average two-time correlation function, usually the average overlap
$C(t)$, or the self part $F_s(\vec{k},t)$ of the intermediate
scattering function~\cite{Ediger2000,Angell2000,Berthier2011}. The
time dependence of the average overlap $C(t)$ for times within one or
two orders of magnitude of the $\alpha$-relaxation time has been found
to be well described by a stretched exponential form $C(t) \sim 
{\rm e}^{-(t/\tau)^{\beta}}$~\cite{Flenner2011}. 

As the relaxation time of a fragile glass former increases in the
vicinity of the glass transition, {\em dynamical heterogeneity\/}
emerges, i.e. the relaxation becomes much slower or much faster in
some regions than in
others~\cite{Ediger2000,Angell2000,Berthier2011a}.  The typical
distance over which the local relaxation is correlated increases as
the glass transition is approached, which, together with other
evidence~\cite{Ediger2000,Angell2000,Berthier2011a}, suggests that
glassy dynamics is a collective phenomenon~\cite{Berthier2011}. One of
the most common approaches to study those correlations is to compute
the four-point structure factor $S_4(\vec{q},t)$, which is the Fourier
transformed spatial correlator of the individual particle overlap
$w$~\cite{Dasgupta1991,Lacevic2003,Berthier2011}.
The dynamic susceptibility $\chi_4(t) \equiv \lim_{q\rightarrow 0}
S_4(\vec{q},t)$~\cite{Lacevic2003,Toninelli2005,Parsaeian2008,Flenner2011,Berthier2011}
gives a measure of the overall strength of the fluctuations, and its
maximum value is sometimes interpreted as being proportional to the
typical number of particles in a correlated region~\cite{Berthier2011}. 
Additionally, a dynamic correlation length $\xi_4(t)$ can be
defined by the expansion $S_4(\vec{q},t) = \chi_4(t) [1 -
  \xi_4^2(t) q^2 + {\mathcal O}(q^4)]$, valid for small but nonzero
$q$~\cite{Lacevic2003,Toninelli2005,Parsaeian2008,Flenner2011}.

The dynamical behavior of glassy systems is characterized by several
timescales besides the relaxation time $\tau_{\alpha}$. Some of them,
such as the time $t_4$ when $\chi_4(t)$ reaches its maximum, are
typically not far from
$\tau_{\alpha}$~\cite{Lacevic2003,Toninelli2005,Berthier2011}. But other
timescales may sometimes be much longer. For example, the dynamic
correlation length of fluctuations continues to increase after
$\tau_{\alpha}$~\cite{Berthier2011,Flenner2011}; and the typical time it
takes for a slow region to become fast or viceversa - the exchange time
$\tau_{\rm ex}$ - may in some cases be much longer than $\tau_{\alpha}$
\cite{Ediger2000,Richert2015,Paeng2015}. However, for times $t > t_4$,
little is known theoretically about $\chi_4(t)$ beyond the observed fact
that it decreases with time~\cite{Toninelli2005,Berthier2011}. In the
case of $\xi_4(t)$, it is not even clear whether it
decreases or not at very long times. Additionally, even though the four
point functions $S_4(\vec{q},t)$ and $\chi_4(t)$ have been the
main tool used to analyze numerical data on dynamical heterogeneity, no
clear connection has been established between them and the exchange time
$\tau_{\rm ex}$ characterizing the lifetime of the heterogeneous
regions. 

The purpose of this work is twofold. First we present evidence that the
long time behavior of the average overlap is given by a power law $C(t)
\sim t^{-d/2}$, where $d$ is the dimensionality of space. Then we focus
on the four point structure factor $S_4(\vec{q},t)$. We introduce a
decomposition of $S_4(\vec{q},t)$ as the sum of four contributions: (i)
$S_4^{\rm cr}$, describing collective relaxation fluctuations; (ii)
$S_4^{\rm st}$, associated with the density fluctuations in the initial
state; (iii) $S_4^{\rm mc}$, due to the interplay of density
fluctuations in the initial state with relaxation fluctuations, and (iv)
$S_4^{\rm sp}$, due to uncorrelated {\em single-particle\/}
fluctuations.  As a function of $q$, $S_4(\vec{q},t)$ decays to a
plateau value $\chi_{4,b} > 0$~\cite{Toninelli2005} for $q \ll
2\pi/r_{NN}$, where $r_{NN}$ is the typical nearest neighbor
distance. We propose here that this plateau corresponds to the sum of
the contributions $S_4^{\rm st}$, $S_4^{\rm mc}$, and $S_4^{\rm sp}$,
and show that its time dependence can be well reproduced by a simple
expression involving only the overlap $C(t)$ and the static structure
factor $S(\vec{q})$. The fact that all contributions except $S_4^{\rm
  cr}$ are $q$-independent allows us to introduce a simple method of
analysis that separates those contributions, and enables the detailed
study of dynamical heterogeneities at timescales much longer than
$\tau_{\alpha}$. We apply this method to simulation data for a binary
hard-sphere system and for the Kob-Andersen Lennard-Jones system. We
show that for temperatures or densities near the mode-coupling
crossover~\cite{Berthier2011,Coslovich2019}, the collective relaxation
contribution $S_4^{\rm cr}$ is orders of magnitude larger than the
others at $t \sim \tau_{\alpha}$, but the single particle contribution
becomes dominant at $t \gg \tau_{\alpha}$. In fact, we find that for
very long times, $S_4(\vec{q},t) \approx S_4^{\rm sp}(\vec{q},t) \approx
C(t) \sim t^{-d/2}$.  By subtracting the other three contributions, we
isolate the collective contribution $S_4^{\rm cr}(\vec{q},t)$, and find
that for the systems we simulate it decays as a power law at very long
times $S_4^{\rm cr}(\vec{q},t) \sim t^{-p}$, with $p > d$, i.e. with an
exponent at least twice larger than the one for the single-particle
contribution.  We also use this decomposition to determine $\xi_4(t)$
for times up to $t \sim 80 \tau_{\alpha}$ in the hard-sphere system,
thus showing how to provide an answer to the longstanding question
regarding the long time behavior of $\xi_4(t)$.

The rest of this paper is organized as follows. In
Sec.~\ref{sec:simulation-details} we briefly discuss the simulation
details. In Sec.~\ref{sec:single-particle} we show evidence for a power
law behavior $C(t) \sim t^{-d/2}$ of the overlap at very long times. In
Sec.~\ref{sec:four-point-functions} we discuss the decomposition of the
four-point functions in terms of four contributions with distinct
physical interpretations, and analyze the long time behavior of the
single-particle and collective relaxation contributions. Finally, in
Sec.~\ref{sec:summary} we summarize our results.

\section{Simulation Details}
\label{sec:simulation-details}
We simulate two 3D equilibrium glass-forming liquids. The first system
is a 50:50 binary mixture of hard-spheres (HARD), with diameters $d$
and $1.4d$. Lengths are measured in units of $d$, and wavevectors are
measured in units of $1/d$. Monte Carlo simulations were performed for
$N=80000$ particles at packing fractions $\varphi =
0.50,0.52,0.55,0.56,0.57$, and $0.58$. For each packing fraction, data
were taken for four runs, after the system was well equilibrated,
during a time of about $100\tau_{\alpha}$. The second system is the
Kob-Andersen Lennard-Jones (KALJ)~\cite{Kob1994, Kob1995, Kob1995a}
80:20 binary mixture with $N=27000$ particles. Here all lengths are
measured in units of $\sigma_{AA}$, the characteristic length of the
Lennard-Jones potential between A particles, and all wavevectors are
measured in units of $1/\sigma_{AA}$. The simulations were performed
with Newtonian dynamics for temperatures $T= 0.50, 0.55, 0.60, 0.65,
0.70$, and $0.80$ at a density $\rho=N/V=1.2$.  At all temperatures,
four runs were performed and data were taken for at least $100
\tau_{\alpha}$ after the system was well equilibrated. More details
about the simulation and characterization of the systems can be found
in Ref.~\cite{Flenner2011} for HARD and in Ref.~\cite{Flenner2014} for
KALJ.

\section{Single-Particle Dynamics}
\label{sec:single-particle}

\begin{figure}[ht!]
  \centering
  \includegraphics[width=\columnwidth] {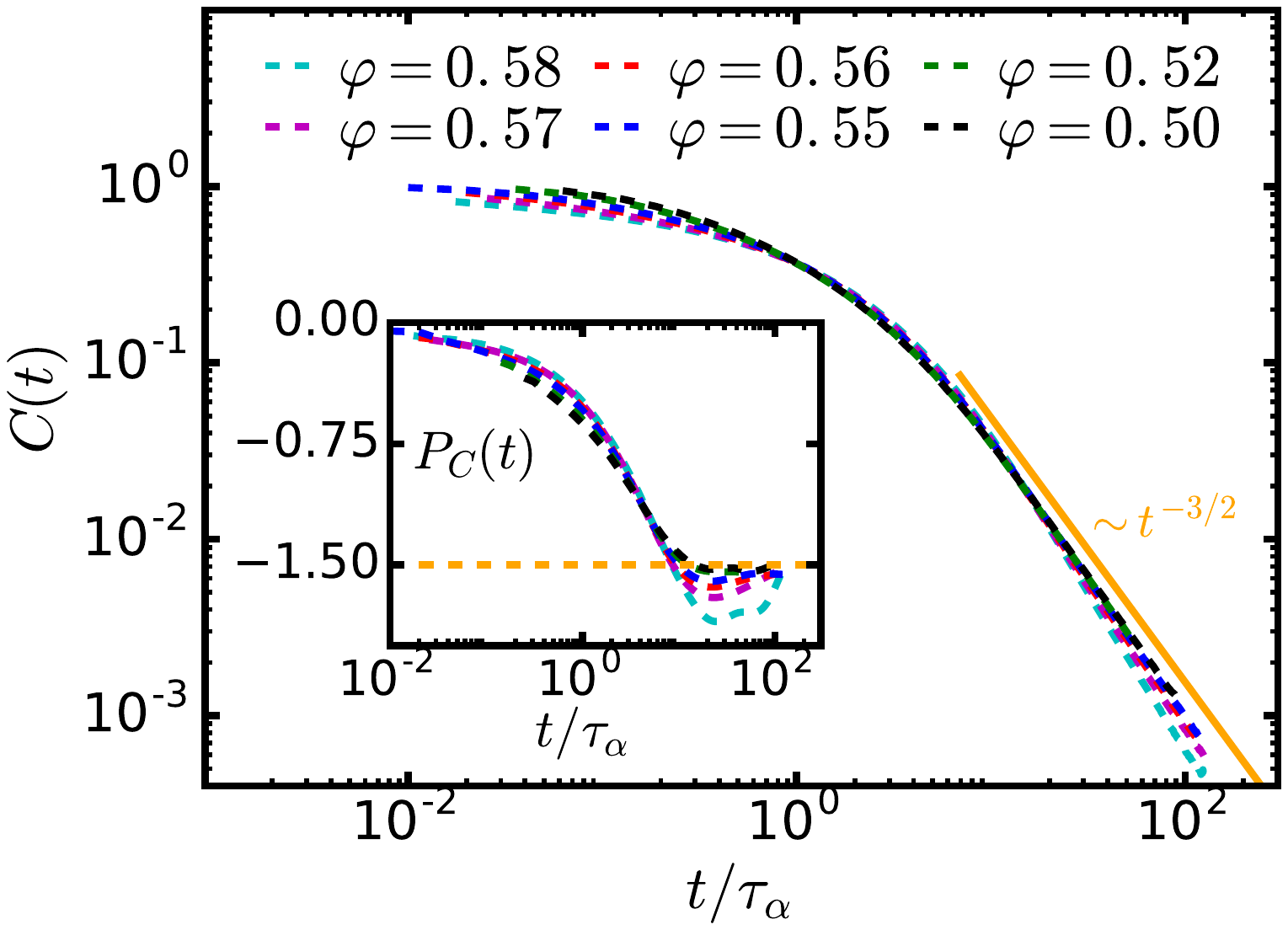}\\[-10.0ex]
  \includegraphics[width=\columnwidth] {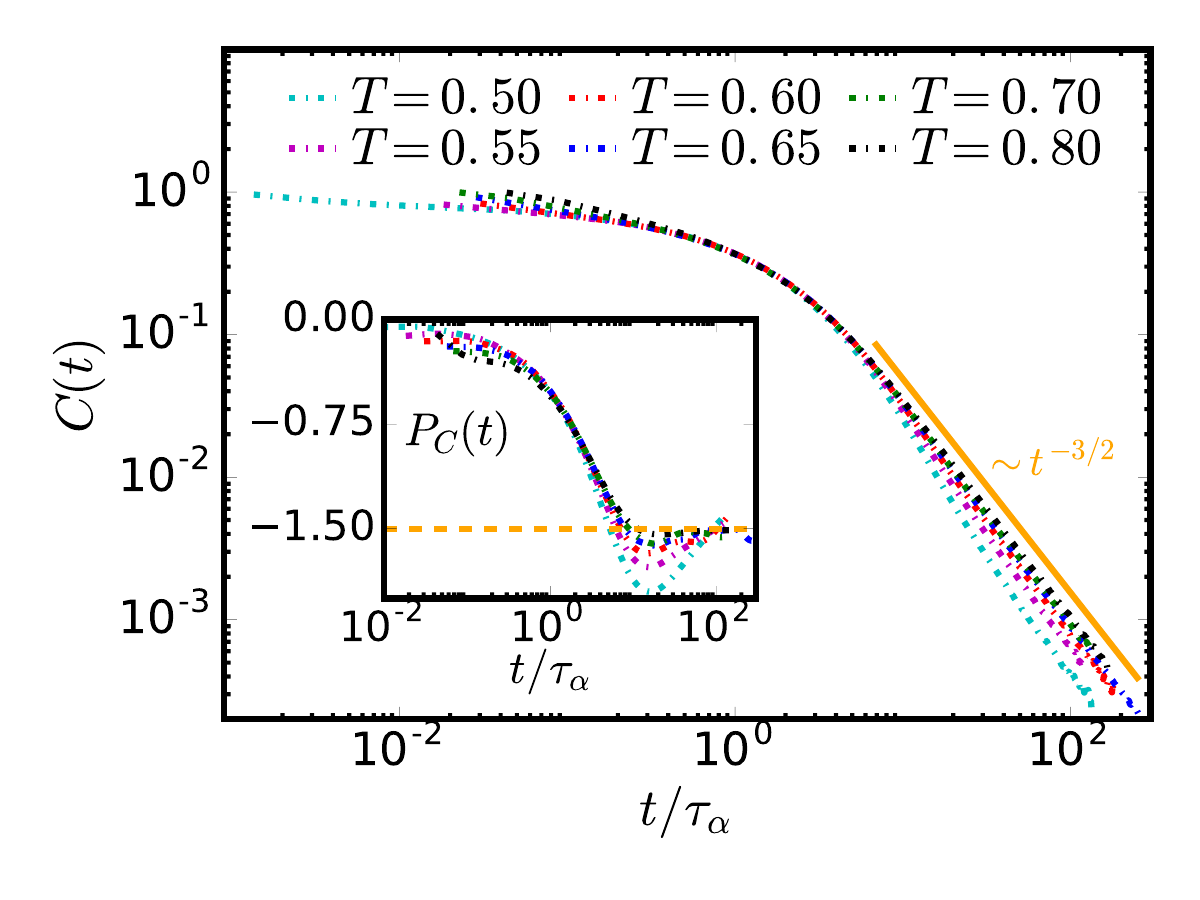}\\[-4.0ex]
  \caption{Two time correlation function $C(t)$ for HARD for packing
    fractions $\varphi=0.50,0.52,0.55,0.56,0.57,0.58$ (top panel) and for
    KALJ for temperatures $T=0.80,0.70,0.65,0.60,0.55,0.50$ (bottom
    panel). A $t^{-3/2}$ power-law time dependence is shown for
    comparison. Insets: Power-law exponent $P_C(t) \equiv d{\ln
      C(t)}/d{\ln t}$ of $C(t)$.  }\label{fig:c_t}
\end{figure}

We probe the dynamics by using a microscopic overlap function
$w_n(t)=\theta [a-|\mathbf{r}_n(t)-\mathbf{r}_n(0)|]$, where $\theta(x)$
is the Heaviside step function, $\mathbf{r}_n(t)$ is the position of the
$n$-$th$ particle at time $t$, and $a$ is a characteristic distance that
is larger than the typical amplitude of vibrational motion (we take $a$
= 0.3 for HARD and $a$ = 0.25 for KALJ). For a given time interval $t$,
if a particle moves less than the characteristic distance $a$, then
$w_n(t)=1$. The average dynamics is characterized by the two-time
correlation $C(t)= N^{-1} \sum_{n=1}^N \langle w_n(t) \rangle $
(i.e. the average fraction of particles with displacements $|\Delta
\vec{r}| < a$), where $\langle ...\rangle$ denotes the average over the
simulation ensemble \cite{Lacevic2003}. We define the
$\alpha$-relaxation time $\tau_{\alpha}$ by setting
$C(\tau_{\alpha})=1/e$.
At times of order $\tau_{\alpha}$, the decay of $C(t)$ follows a
stretched exponential form $C(t) \sim
\exp[-(t/\tau)^{\beta}]$~\footnote{For HARD, there is a weakly $\varphi$
  dependent exponent $\beta \approx 0.55$~\cite{Flenner2011}. For KALJ,
  the exponent is in the range $0.5 \alt \beta \alt 0.7$ for the range
  of temperatures discussed in this work.}.

However, as shown in Fig.~\ref{fig:c_t}, at times $t \gg \tau_{\alpha}$
the decay of $C(t)$ approaches a power law form, both for the HARD and
the KALJ systems. The insets of Fig.~\ref{fig:c_t} show that the exponent
$P_C(t) \equiv d{\ln C(t)}/d{\ln t}$ approaches $-1.5 = -d/2$ at very
long times, where $d=3$ is the dimensionality. A fuller discussion of
this limit is given in~\cite{Castillo2020}, but we can give a simple
argument to justify this behavior. At time $t \gg \tau_{\alpha}$, and
considering only long lengthscales, we expect the dynamics to be
diffusive with self-diffusion coefficient $D$, and the diplacement
probability distribution $G_s(\vec{r},t)$ to be a gaussian with
characteristic size $R(t) = (2 D t)^{1/2}$. This corresponds to
$G_s(\vec{0},t) \approx (4\pi D t)^{-d/2}$, and the probability of being
within a region of radius $a \ll R(t)$ in dimension $d$ around the
origin to be $C(t) \sim a^d G_s(\vec{0},t) \sim
t^{-d/2}$~\cite{Castillo2020}.
We expect $G_s(\vec{r},t)$ to approach normal-diffusion-like behavior
and $P_C(t)$ to approach $-d/2$ faster for less glassy systems (more
weakly interacting, lower $\varphi$, higher $T$) and viceversa. Indeed,
$P_C(t)$ ``overshoots'' its asymptotic value of $-d/2$ for $10
\tau_{\alpha} \alt t < 100 \tau_{\alpha}$, and this overshooting
increases with $\varphi$ for HARD (Fig.~\ref{fig:c_t}, top panel inset) and
increases at lower $T$ for KALJ (Fig.~\ref{fig:c_t}, bottom panel inset).

\section{Four-Point Functions: Decomposition and Long-time behavior}
\label{sec:four-point-functions}

\subsection{Contributions to the Four-Point Functions}
\label{sec:four-point-function-decomposition}

To characterize the dynamical heterogeneity, we
compute~\cite{OhioSupercomputerCenter1987} the four-point dynamic
structure factor $S_4(\vec{q},t)$~\cite{Dasgupta1991,Lacevic2003},
\begin{equation}
  \label{eq:S-4}
  S_4(\vec{q},t) \equiv \frac{1}{N} \! \sum_{n,n^{\prime}=1}^{N} \! \!
  \la w_n(t)  w_{n^\prime}(t)
  e^{[i\vec{q}\cdot( \vec{r}_n(0) -  \vec{r}_{n^\prime}(0) )]} \ra
  -\delta_{\vec{q},0}N C^2(t)
\end{equation}

\begin{figure}[ht!]
  \centering
  \includegraphics[width=\columnwidth]{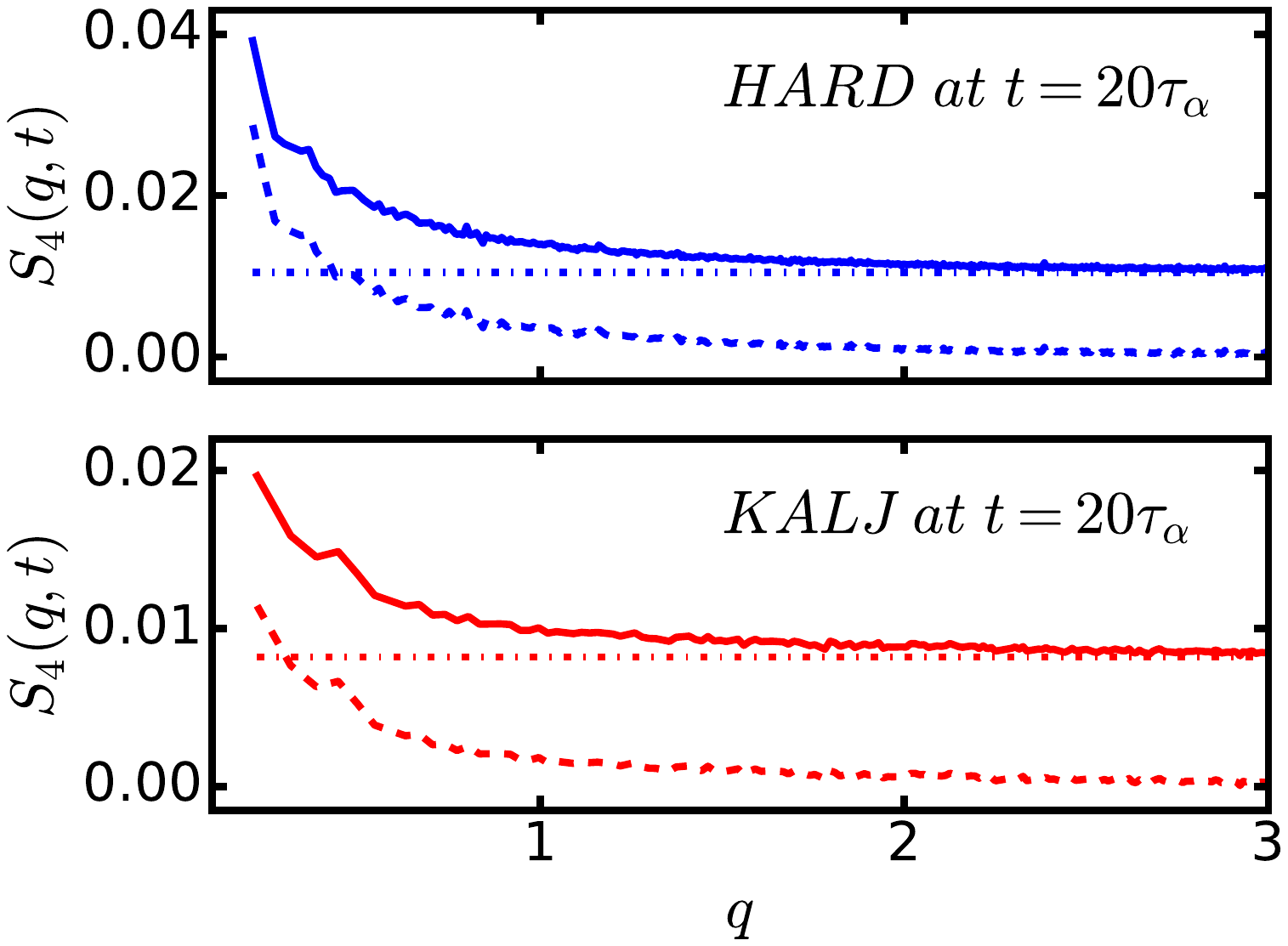}\\[-3.0ex]
  \includegraphics[width=\columnwidth]{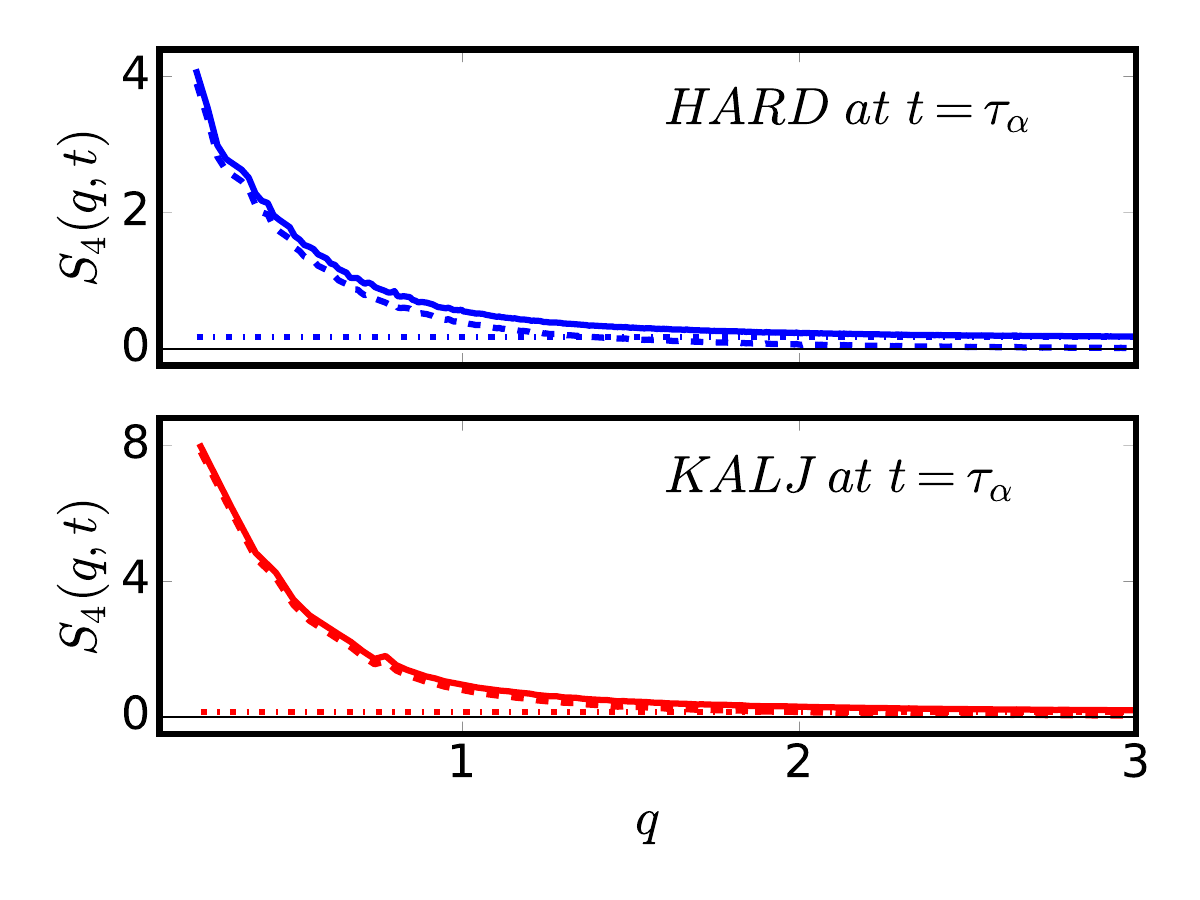}\\[-4.0ex]
  \caption{Decomposition of $S_4(\vec{q},t)$ (full lines): collective
    relaxation part $S_4^{\rm cr}(\vec{q},t)$ (dashed lines), and
    background term $\chi_{4,b}(t)$ (dashed-dotted lines). First panel
    from top: HARD at $\varphi=0.57$, with $t = 20 \tau_{\alpha}$. Second
    panel: KALJ at $T=0.55$, with $t = 20 \tau_{\alpha}$. Third panel: HARD at
    $\varphi=0.57$, with $t = \tau_{\alpha}$. Bottom panel: KALJ at
    $T=0.55$, with $t = \tau_{\alpha}$.
  }
  \label{fig:decomposition_s_4}
\end{figure}

The full lines in Fig.~\ref{fig:decomposition_s_4} show
$S_4(\vec{q},t)$. The two top panels correspond to time around $20
\tau_{\alpha}$ for both systems. Let's consider intermediate values of
$q$, $\xi_4^{-1} \ll q \ll q_0 \approx 2\pi/r_{NN}$, where $q_0$ is the
location of the main peak of the static structure factor $S(q)$, and
$r_{NN}$ is the typical nearest neighbor distance. We find that at $t
\approx 20 \tau_{\alpha}$, for $\xi_4^{-1} \ll q \ll q_0$,
$S_4(\vec{q},t)$ decays to a plateau value $\chi_{4,b}$ which is almost
half of its maximum $\chi_4$ at the origin.
By contrast, at $t \approx \tau_{\alpha}$, the $q$-independent
background $\chi_{4,b}$ is very small compared to the peak value
$\chi_{4}$, as shown in the bottom two panels of the same figure. 
A $q$-independent background in Fourier space suggests that
there are uncorrelated displacements of particles in position
space, giving rise to $S_4^{\rm sp}(\vec{q},t) \ne 0$. As discussed in
~\cite{Castillo2020}, $S_4^{\rm sp}(\vec{q},t)$ only contains
contributions from same particle ($n=n'$) terms in Eq.~(\ref{eq:S-4}),
and by neglecting a small collective relaxation contribution to those terms we
obtain
\begin{equation}
  \label{eq:S4sp-C-C2}
  S_4^{\rm sp}(\vec{q},t) = \chi_4^{\rm sp}(t) \approx \frac{1}{N}
  \sum_{n=1}^{N} \langle (w_n(t) -\langle w_n(t) \rangle)^2 \rangle
  = C(t)-C^2(t).
\end{equation}

\begin{figure}[ht]
  \centering
  \includegraphics[width=\columnwidth] {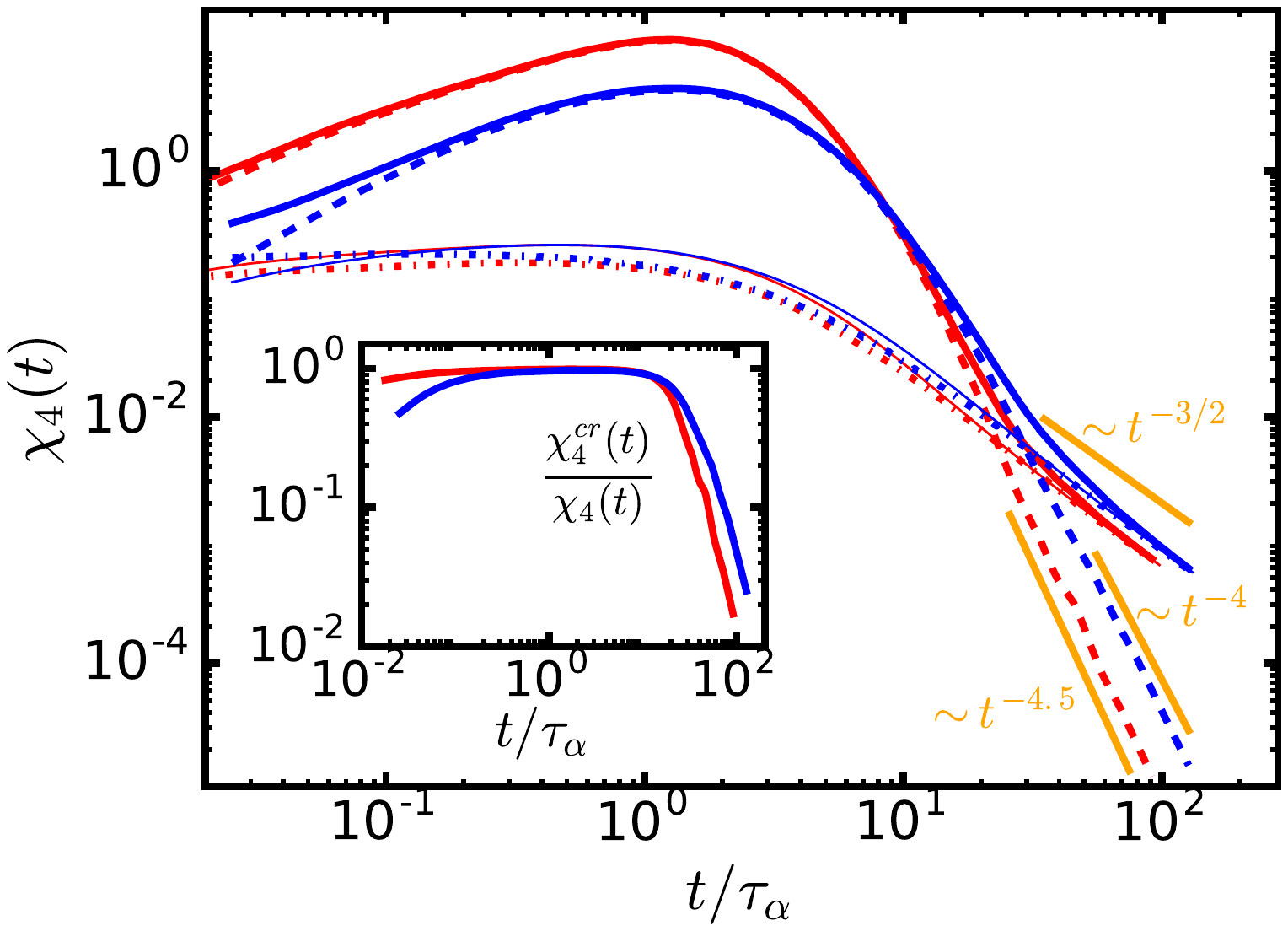}\\[-4.0ex]
  \caption{Dynamic susceptibility decomposition for HARD at $\varphi=0.57$
   (blue) and KALJ at $T=0.55$ (red). Main panel: total dynamic
   susceptibility $\chi_4(t)$ (solid lines), collective relaxation
   part $\chi^{\rm cr}_4(t)$ (dashed lines), background term
   $\chi_{4,b}(t)$ (dashed-dotted lines), and its leading order
   approximation $\chi_{4,b}^{(0)}(t)$ (thin full lines). Power law
   time dependences (orange full lines) shown for comparison with long
   time asymptotic behavior for: $\chi_{4,b}(t)$ ($\sim t^{-3/2}$),
   $\chi^{\rm cr}_4(t)$ for HARD ($\sim t^{-4} \ll t^{-3}$), and
   $\chi^{\rm cr}_4(t)$ for KALJ ($\sim t^{-4.5} \ll t^{-3}$). Inset:
   ${\chi^{\rm cr}_4(t)}/{\chi_4(t)}$.}\label{fig:decomposition_chi_4}
\end{figure}

The initial density contribution $S_4^{\rm st}(\vec{q},t)$ is
obtained~\cite{Castillo2020} by replacing the microscopic overlap
$w_n(t)$ by its average $C(t) = \langle w_n(t) \rangle$ in
Eq.~(\ref{eq:S-4}),
\begin{eqnarray}
  \label{eq:S-4st}
  S_4^{\rm st}(\vec{q},t) \! \! \! \! & \equiv & \! \! \! \! 
  \frac{1}{N} \! \sum_{n,n^{\prime}=1}^{N} \! \!  C^2(t)
  \langle e^{[i\vec{q}\cdot( \vec{r}_n(0) -  \vec{r}_{n^\prime}(0) )]} \rangle
  -\delta_{\vec{q},0}N C^2(t) \nonumber \\
  & = & C^2(t) [S(\vec{q}) -\delta_{\vec{q},0}N].
\end{eqnarray}

For $q \ll q_0 \approx 2\pi/r_{NN}$, the static structure factor is
weakly dependent on $q$, and
$S(q) \approx \lim_{q \to 0} S(q) = N^{-1} \langle (\delta N)^2
\rangle$. In Reference~\cite{Castillo2020} it is argued that, for
$\xi_4^{-1}, q \ll 2\pi/r_{NN}$, the $q$-dependence of all
contributions except $S_4^{\rm cr}$ can be neglected, so that
\begin{eqnarray}
  S_4(\vec{q},t) & \approx & S_4^{\rm cr}(\vec{q},t) + \chi_{4,b}(t),
  \quad \text{with} 
  \label{eq:S4-S4coll-chi4b} \\
  \chi_{4,b}(t) & \equiv & \chi_4^{\rm sp}(t) 
  + \lim_{q \to 0} S_4^{\rm st}(\vec{q},t)
  + \lim_{q \to 0} S_4^{\rm mc}(\vec{q},t)
  \label{eq:chi4b-chisp-chi4st-chi4mc} \\
  & \approx & C(t) + [N^{-1} \langle (\delta N)^2 \rangle - 1] C(t)^2
  \equiv \chi_{4,b}^{(0)}(t).
  \label{eq:chi4b0-def}
\end{eqnarray}

To extract the collective relaxation part $S_4^{\rm cr}(\vec{q},t)$
from the data, we use Eq.~(\ref{eq:S4-S4coll-chi4b}). 
Fig.~\ref{fig:decomposition_s_4} shows this decomposition:
$S_4(\vec{q},t)$ is shown with full lines, the $q$-independent
background $\chi_{4,b}(t)$ is shown with dash-dotted lines, and
$S_4^{\rm cr}(\vec{q},t)$ is shown with dashed lines. 

To characterize the collective relaxation part of the four-point
function, $S_4^{\rm cr}(\vec{q},t)$, we fitted it with a slightly
generalized version of the Ornstein-Zernike functional form, motivated
by results from inhomogeneous mode coupling theory~\cite{Biroli2006},
which has been used in~\cite{Karmakar2010, Flenner2011, Flenner2010}
(see App.~\ref{sec:FittingMethod} for more details on the fitting
procedure). The fitting form for $S_4^{\rm
  cr}(\vec{q},t)$ reads
\begin{equation}
  \label{eq:modified-oz}
  S_4^{\rm cr}(\vec{q},t)=\frac{\chi_4^{\rm cr}(t)}{1+ [\xi_4^{\rm cr}(t)]^2 q^2
   + [c(t)]^2 q^4}, 
\end{equation} 
where $\chi_4^{\rm cr}(t)$ is the collective relaxation part of the
dynamic susceptibility, $\xi_4^{\rm cr}(t)$ is the four-point
dynamic correlation length, and $c(t)$ is an additional parameter
characterizing the quartic contribution. From this point
on we use the notation $\xi_4^{\rm cr}(t)$ for this correlation
length, to emphasize that it is extracted from the collective
relaxation part of the four-point function.

The presence of additional contributions beyond the one due to
collective relaxation cannot be ignored, particularly for long times $ t
\agt 10 \tau_{\alpha}$. It is shown in Appendix~\ref{sec:compare-fits}
that attempting to fit the data in that time regime without taking into
account those additional contributions leads to very poor fits and to
substantial systematic errors in the determination of the dynamic
susceptibility $\chi_4(t)$ and the dynamic correlation length
$\xi_4^{\rm cr}(t)$.

\subsection{Long-time behavior of the dynamic susceptibility $\chi_4(t)$
  and the single particle and collective relaxation contributions}
\label{sec:long-time-chi4}

By taking the $q \to 0$ limit of Eq.~(\ref{eq:S4-S4coll-chi4b}), we
obtain the decomposition $\chi_4(t) \approx \chi_4^{\rm cr}(t) +
\chi_{4,b}(t)$ for the dynamic susceptibility. 
Fig.~\ref{fig:decomposition_chi_4} shows $\chi_4(t)$ (full
lines), $\chi_4^{\rm cr}(t)$ (dashed lines), $\chi_{4,b}(t)$
(dash-dotted lines), and $\chi_{4,b}^{(0)}(t)$ (thin full lines), in the cases
of HARD for packing fraction $\varphi=0.57$ (blue) and KALJ for temperature
$T=0.55$ (red).
For both systems, the collective relaxation part $\chi_4^{\rm cr}(t)$ of
the dynamic susceptibility increases with time to a peak value
$\chi_4^{\rm cr, max}$, which
may be interpreted to indicate the maximum correlated volume of the
fluctuating region.
The approximation $\chi_{4,b}(t) \approx \chi_{4,b}^{(0)}(t)$
(Eq.~(\ref{eq:chi4b0-def})) becomes asymptotically exact
for $t \gg \tau_{\alpha}$, and the biggest discrepancy between the two
quantities is $\chi_{4,b}(t) / \chi_{4,b}^{(0)}(t) \approx 0.7$ when
$\chi_4^{\rm cr}(t)$ is near its peak, i.e. when the collective
relaxation corrections neglected in
Eq.~(\ref{eq:chi4b0-def}) are largest~\cite{Castillo2020}.
For long times, $t \gg \tau_{\alpha}$, $\chi_4^{\rm cr}(t)$ decreases
as a $t^{-3}$ power law or faster, while $\chi_{4,b}(t)$ - which in
this time regime is dominated by $\chi_4^{\rm sp}(t) \approx C(t)$ -
also decreases but as a much slower power law $\sim
t^{-3/2}$~\cite{Castillo2020}. Thus there is a crossover between a
shorter time regime where the collective relaxation contribution
dominates and a longer time regime where the single particle
contribution dominates. We define the crossover time
$\tau_{\chi_4^{\rm cr}/\chi_4=1/2}$ as the time when $\chi_4^{\rm
  cr}(t)/\chi_4(t) = 1/2$. We find that $\tau_{\chi_4^{\rm
    cr}/\chi_4=1/2} \sim 40 \tau_{\alpha}$ for HARD at $\varphi=0.57$ and
$\tau_{\chi_4^{\rm cr}/\chi_4=1/2} \sim 25 \tau_{\alpha}$ for KALJ at
$T=0.55$.
The inset of Fig.~\ref{fig:decomposition_chi_4} shows the ratio
${\chi^{\rm cr}_4(t)}/{\chi_4(t)}$ for the same cases as in the main
panel. The ratio is close to unity for times up to about
$20\tau_{\alpha}$ and then it decreases rapidly, becoming roughly two
orders of magnitude smaller by $t \sim 100\tau_{\alpha}$. For other
values of the control parameters, as long as the system is close to
the mode-coupling crossover, the behavior of ${\chi^{\rm
    cr}_4(t)}/{\chi_4(t)}$ is very similar~\cite{Pandit2020b}.

\begin{figure}[ht!]
  \centering
  \includegraphics[width=\columnwidth]{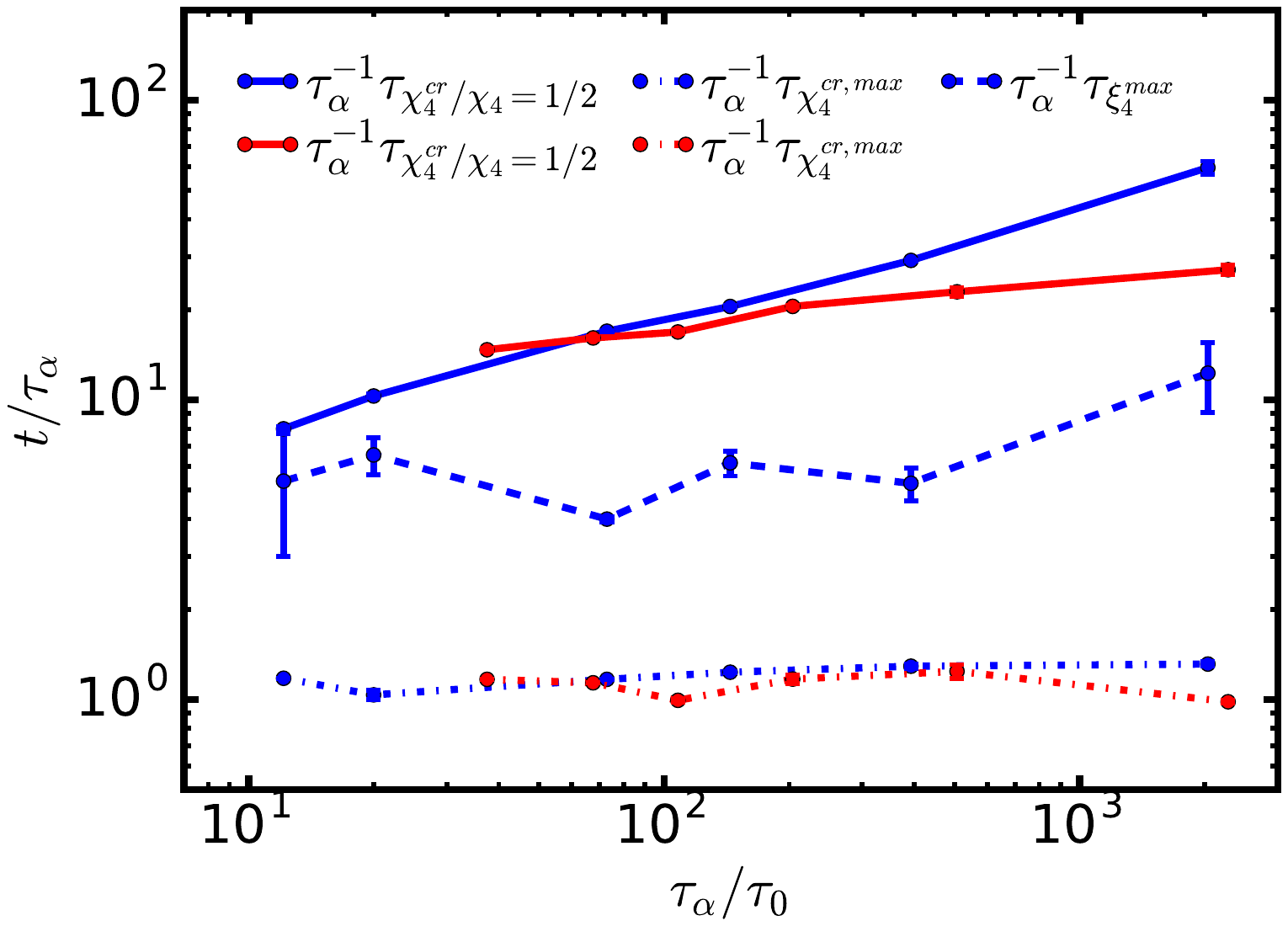}\\[-4.0ex]
  \caption{Rescaled timescales for HARD (blue) and KALJ (red) as
    functions of the rescaled relaxation time
    $\tau_{\alpha}/\tau_0$~(\cite{Flenner2014}): rescaled crossover
    time $\tau_{\alpha}^{-1} \tau_{\chi_4^{\rm cr}/\chi_4=1/2}$ (solid
    lines); rescaled time $\tau_{\alpha}^{-1} \tau_{\xi_{4}^{\rm
        max}}$ at which the dynamic correlation length $\xi_4^{\rm
      cr}$ becomes maximum (dashed lines); rescaled time
    $\tau_{\alpha}^{-1} \tau_{\chi_{4}^{\rm cr, max}}$ at which the
    collective dynamic susceptibility $\chi_4^{\rm cr}$ becomes
    maximum (dash-dotted lines). Following Ref.~\cite{Flenner2014},
    the parameter $\tau_0$ ($\tau_0=70$ for HARD and $\tau_0=1/15$ for
    KALJ) is used so that relaxation times can be compared across different
    systems.}\label{fig:time-scales}
\end{figure}

As the system approaches the glass transition at fixed rescaled time
$t/\tau_{\alpha}$, the collective relaxation contribution $\chi_4^{\rm
  cr}$ grows strongly, while the background contribution $\chi_{4,b}$,
which, to a good approximation, can be computed in terms of $C(t)$ and
$S(q)$ (Eq.~(\ref{eq:chi4b0-def})), shows little if any
change~\cite{Castillo2020}. Thus we expect both the ratio ${\chi^{\rm
    cr}_4(t)}/{\chi_4(t)}$ at fixed $t/\tau_{\alpha}$~\cite{Pandit2020b}
and the rescaled crossover time ${\tau_{\alpha}}^{-1}
\tau_{\chi_4^{\rm cr}/\chi_4=1/2}$ (Fig.~\ref{fig:time-scales}) to
increase. Both increases are indeed observed in our data, and in fact
we find $\tau_{\chi_4^{\rm cr}/\chi_4=1/2} \sim \tau_{\alpha}^{1+p}$,
with $p \approx 0.40$ for HARD and $p \approx 0.15$ for KALJ.

\begin{figure}[ht!]
  \centering
  \includegraphics[width=\columnwidth]{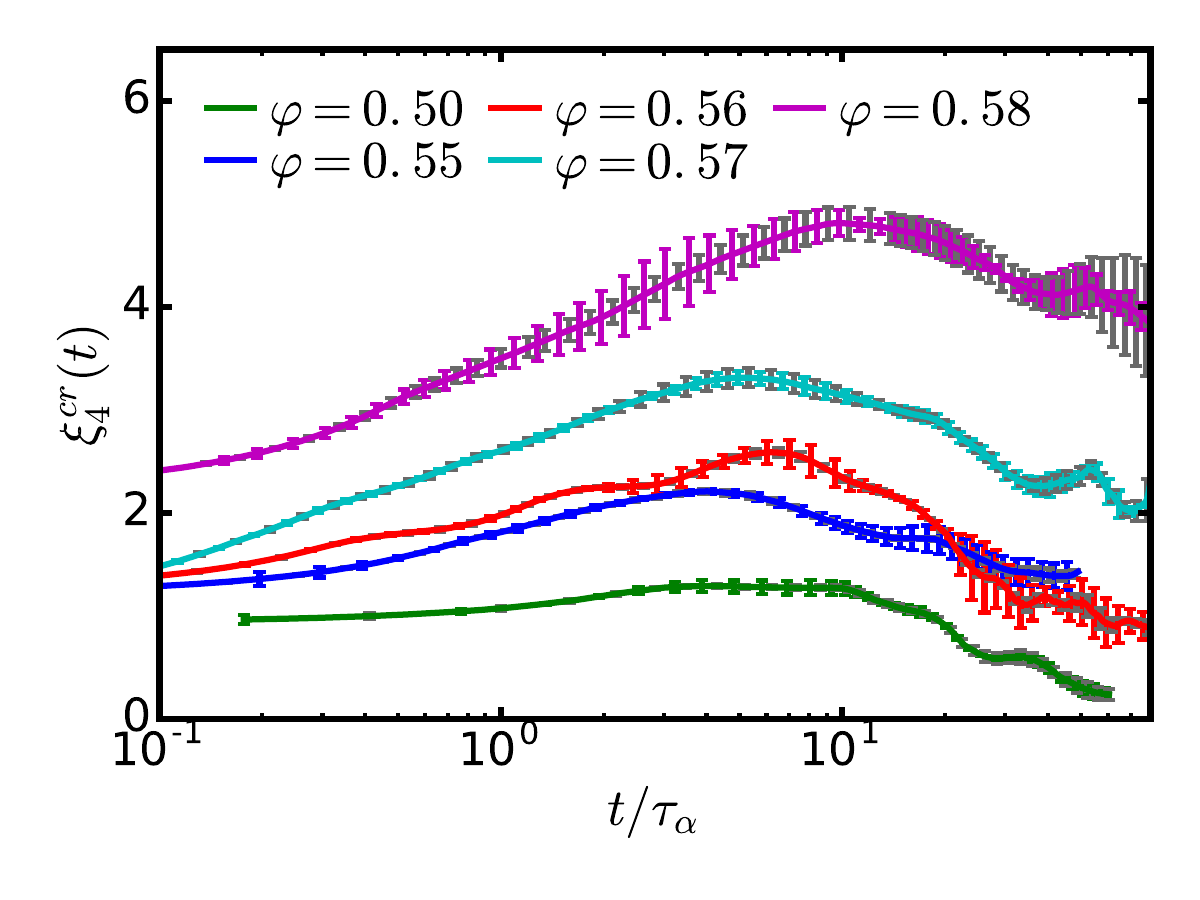}\\[-4.0ex]
  \caption{Correlation length $\xi_4^{\rm cr}(t)$ as a function of
    $t/\tau_{\alpha}$ for HARD system at packing fractions
    $\varphi=0.50,0.55,0.56,0.57,0.58$. The grey error bars represent
    statistical errors in the determination of $\xi_4^{\rm cr}(t)$ for a
    fixed fitting interval. The error bars in the same color as the
    curve represent the range of variation in the determination of
    $\xi_4^{\rm cr}(t)$ as the fitting interval is changed (see
    App.~\ref{sec:FittingMethod}). For clarity, each type of error bar
    is showed for one out of every 10 data points.}\label{fig:xi_t}
\end{figure}

\subsection{Long-time behavior of the correlation length $\xi_4(t)$
  for the binary hard-sphere system}
\label{sec:long-time-xi4}

We now turn to the determination of the dynamic correlation
length.
The behavior of the dynamic correlation length $\xi_4(t)$ in
glass-forming liquids for times $t > \tau_{\alpha}$ has been
controversial. 
In one early study~\cite{Lacevic2003}, it was found that the
time dependence of the dynamic correlation length roughly follows that
of the dynamic susceptibility. Other studies, in a variety of
glass-forming models, have found monotonous increasing growth of the
dynamic correlation length as time increases~\cite{Toninelli2005},
possibly with a plateau~\cite{Doliwa2000,Rotman2010} starting at a
time longer than both $\tau_{\alpha}$ and the time when $\chi_4$
reaches its peak. Monotonous growth of the dynamic correlation length
with time difference was also found in aging glassy
systems~\cite{Parsaeian2008,Parsaeian2008a}.
In Fig.~\ref{fig:xi_t},
we show results for $\xi_4^{\rm cr}(t)$ as a function of
$t/\tau_{\alpha}$ for times up to $80\tau_{\alpha}$ for the HARD
system at packing fractions $\varphi=0.50,0.52,0.55,0.57,0.58$. As
discussed in~\cite{Flenner2011}, the dynamic correlation length grows
approximately logarithmically with times and reaches a maximum value
$\xi_{\rm max} > \xi_4^{\rm cr}(\tau_{\alpha})$ at a time
$\tau_{\xi_{4}^{\rm max}} > \tau_{\alpha}$. The approach introduced in
this work allows us to now explore times $t \gg \tau_{\xi_{4}^{\rm
    max}}$. We find that for $\tau_{\rm max} < t \alt 80\tau_{\alpha}$,
our results for $\xi_4^{\rm cr}(t)$ are noisy, but they show a general
trend to decrease as time increases. 

\section{Summary}
\label{sec:summary}

In this paper we have discussed the behavior of the average overlap and
of the four-point functions in models of glass-forming liquids, with
emphasis on times much longer than the $\alpha$-relaxation time. We have
presented simulation results for two models of 3D glass forming liquids:
a binary hard-sphere model and a Kob-Andersen Lennard-Jones model. We
have showed that at very long times the average overlap $C(t)$ probing
the similarity between an initial and a final state separated by a time
interval $t$ decays as a power law $C(t) \sim t^{-d/2}$. This is much
slower than the stretched exponential behavior $C(t) \sim {\rm
  e}^{-(t/\tau)^{\beta}}$ previously observed at times within one or two
  orders of magnitude of the $\alpha$-relaxation time $\tau_{\alpha}$.

We have also introduced a decomposition of the four point dynamic
structure factor $S_4(\vec{q},t)$ as the sum of four parts: $S_4^{\rm
  cr}$ (collective relaxation fluctuations); $S_4^{\rm sp}$ ({\em
  single-particle\/} fluctuations); $S_4^{\rm st}$ (initial density
correlations); and $S_4^{\rm mc}$ (interplay between initial density
fluctuations and collective relaxation fluctuations). Although valid at
all times, this decomposition is particularly useful to enable the study
of dynamical heterogeneities at $t \gg \tau_{\alpha}$.  We argued that
in this decomposition, all contributions except the collective
relaxation one can be approximated as $q$-independent for $q \ll
2\pi/r_{NN}$, thus explaining the presence of a flat background term
$\chi_{4,b}(t)$ in the $q$-dependence of the four-point function, as
made explicit in Eq.~\ref{eq:S4-S4coll-chi4b}. This structure allowed us
to subtract the background from $S_4(\vec{q},t)$ and thus recover the
collective relaxation contribution $S_4^{\rm cr}$. We have also shown
that a simple approximate expression depending only on the overlap
$C(t)$ and the static structure factor $S(\vec{q})$ reproduces very well
the time dependence of the background term, particularly for times $t
\gg \tau_{\alpha}$.

We have found that for higher $\varphi$ (lower $T$), $S_4^{\rm cr}$ is
between one and two orders of magnitude bigger than the other
contributions at $t \sim \tau_{\alpha}$, but for $t \gg \tau_{\alpha}$
the single particle contribution $S_4^{\rm sp} \approx C(t) \propto
t^{-d/2}$ dominates against all others, because $S_4^{\rm cr} + S_4^{\rm
  st} + S_4^{\rm mc} \alt {\rm const} \; t^{-d}$.
We have also used the decomposition of $S_4(\vec{q},t)$ to address the
controversy regarding $\xi_4(t)$ for $t \gg \tau_{\alpha}$: for a binary
hard-sphere mixture, we found that $\xi_4(t)$ is maximum at $t =
\tau_{\xi_{4}^{\rm max}} \sim 4-15 \tau_{\alpha}$ and then generally
decreases up to at least $t \sim 80 \tau_{\alpha}$.

The decomposition introduced here enables substantial further progress
in the understanding of dynamical heterogeneities in glassy systems. A
first application~\cite{Pandit2020c} will introduce an explicit formula
for $S_4^{\rm cr}(\vec{q},t)$ in terms of the average correlation
function $C(t)$ and a two-point correlation function $s(\vec{q},t)$ of
the local relaxation rates. This two-point function $s(\vec{q},t)$
probes the collective relaxation dynamics and makes quantitative the
qualitative description of dynamic heterogeneity in terms of slow and
fast regions. It also provides a method to obtain $\tau_{\rm ex}$ from
$S_4(\vec{q},t)$~\cite{Pandit2020c}, and allows to obtain explicit
predictions for $\chi_4(t)$ under various assumptions regarding the
decay of the relaxation rate fluctuations. Potential applications of the
same ideas also include, among others, the introduction of other
observables that are better able to probe the relaxation rate
fluctuations, and the study of spatiotemporal correlations of local
relaxation rates in aging systems.

\section{Acknowledgement}

R.~K.~P.~acknowledges the Ohio University Condensed Matter and Surface
Sciences (CMSS) program for support through a studentship.

\appendix

\section{Fitting Method}
\label{sec:FittingMethod}

To extract the collective relaxation part of the four-point function,
$S_4^{\rm cr}(\vec{q},t)$ and the $q$-independent background $\chi_{4,b}$, we
fitted $S_4(\vec{q},t)$ by combining Eqs.~(\ref{eq:S4-S4coll-chi4b}) and
(\ref{eq:modified-oz}). 
The complete fitting form for $S_4(\vec{q},t)$ reads
\begin{equation}
  \label{eq:modified-oz-chi4b}
  S_4(\vec{q},t)=\frac{\chi_4^{\rm cr}(t)}{1+ [\xi_4^{\rm cr}(t)]^2 q^2
   + [c(t)]^2 q^4} + \chi_{4,b}(t). 
\end{equation} 

\begin{figure}[ht!]
  \centering
  \includegraphics[width=\columnwidth]{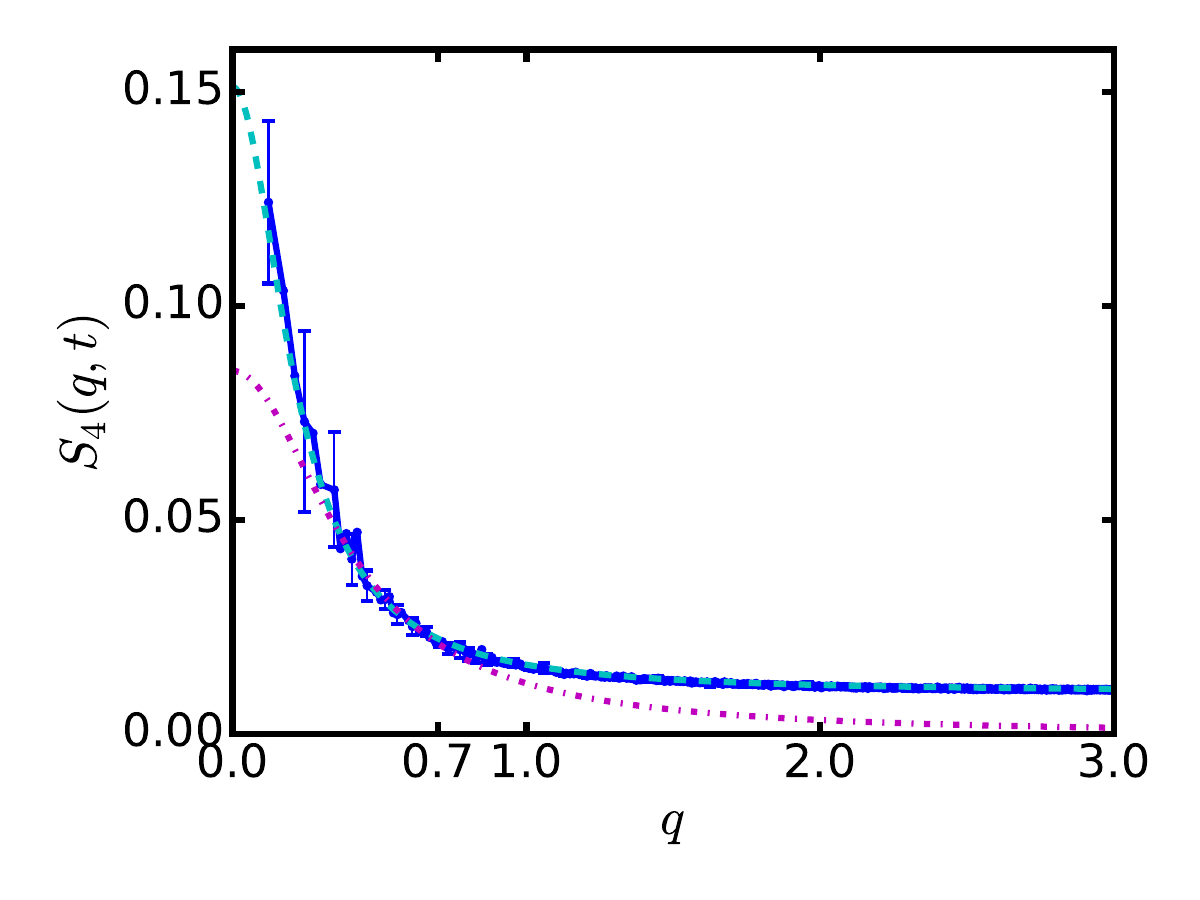}\\[-4.0ex]
  \caption{Effects of allowing for a background contribution to
    $S_4(\vec{q},t)$ due to single particle fluctuations and initial
    density fluctuations. $S_4(\vec{q},t)$ as a function of $q$ for HARD
    at $\varphi=0.58, t = 20 \tau_{\alpha}$: data (symbols with error bars
    joined by full line), fit allowing for a q-independent background
    (dashed line), fit not allowing for a background contribution
    (dot-dashed line).}\label{fig:comparing-fits-vs-q_s_4}
\end{figure}

\begin{figure}[ht!]
  \centering
  \includegraphics[width=\columnwidth]{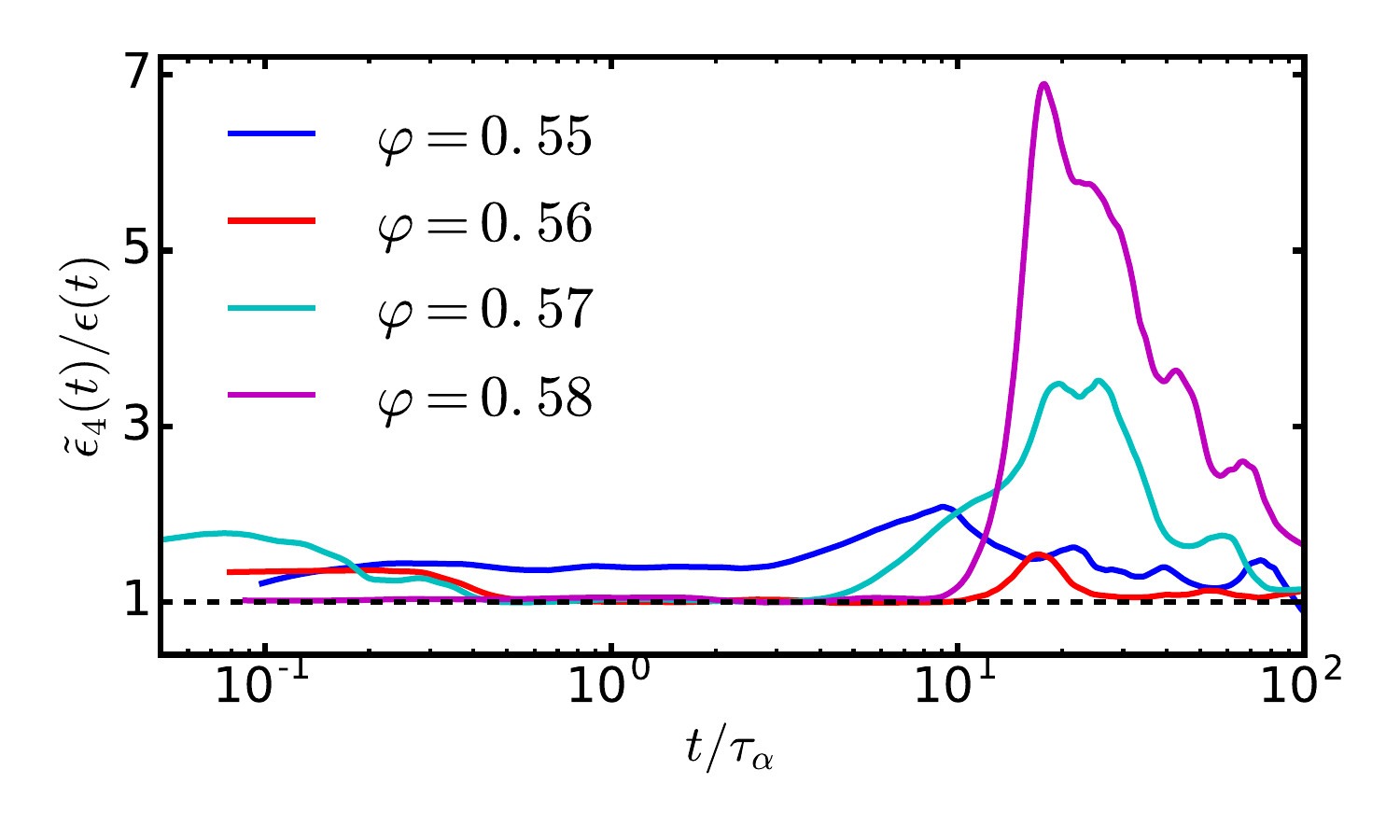}\\[-1.7ex]
  \includegraphics[width=\columnwidth]{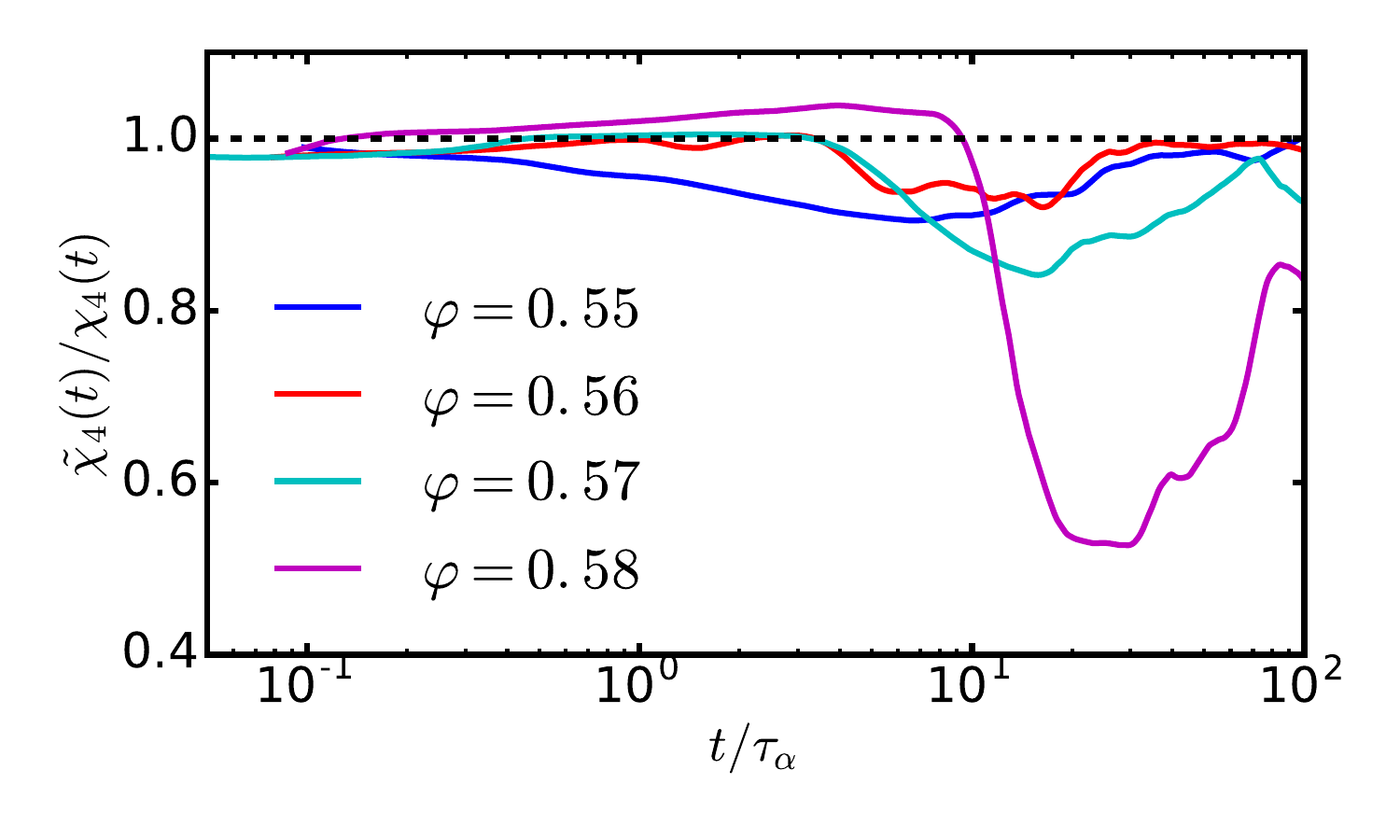}\\[-1.7ex]
  \includegraphics[width=\columnwidth]{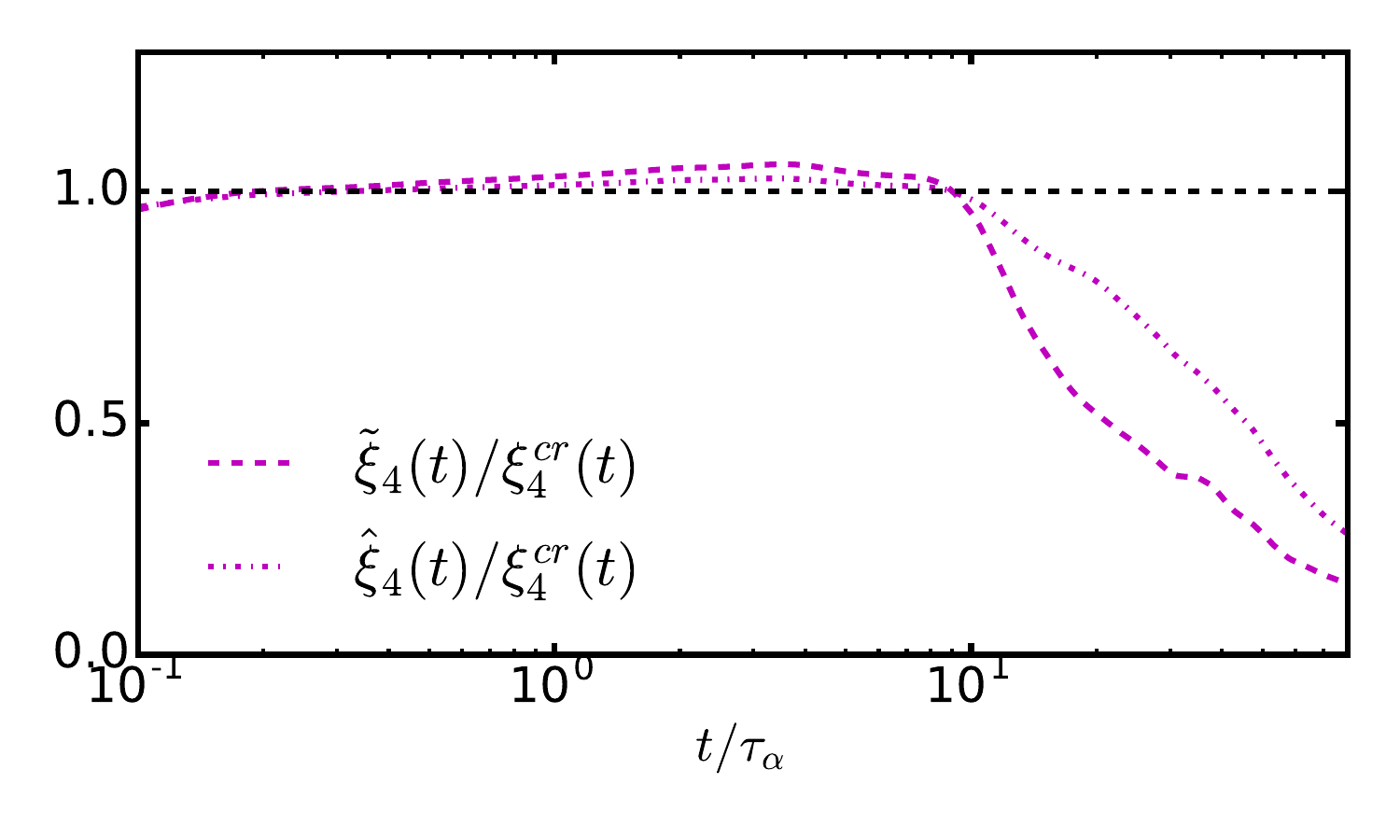}\\[-2.7ex]
 \caption{Effects of allowing for a $q$-independent background
   contribution $\chi_{4,b}(t)$ to $S_4(\vec{q},t)$ due to single
   particle fluctuations and initial density fluctuations, in the case
   of HARD. $\chi_4(t)$ [$\tilde{\chi}_4(t)$] is the dynamic
   susceptibility, $\xi_4^{\rm cr}(t)$ [$\tilde{\xi}_4(t)$] is the dynamic
   correlation length, and $\epsilon(t)$ [$\tilde{\epsilon}(t)$] is the
   rms fitting error per degree of freedom in the interval $0 < q <
   q_{\epsilon}=0.4$, obtained from a fit of $S_4(\vec{q},t)$ vs $q$
   allowing [not allowing] for a background
   contribution. $\hat{\xi}_4(t)$ is the dynamic correlation
   length obtained from the following procedure: first, $\chi_4(t)$ is
   obtained from a fit allowing for a nonzero background; after that, a
   fit is performed where $\chi_4(t)$ is fixed to the value obtained
   before, but the background is constrained to be zero. Top panel:
   $\tilde{\epsilon}(t)/\epsilon(t)$ vs time $t$, for
   $\varphi=0.55,0.56,0.57,0.58$. Middle panel:
   $\tilde{\chi}_4(t)/\chi_4(t)$ vs time $t$, for
   $\varphi=0.55,0.56,0.57,0.58$. Bottom panel:
   $\tilde{\xi}_4(t)/\xi_4^{\rm cr}(t)$ and $\hat{\xi}_4(t)/\xi_4^{\rm cr}(t)$
   vs time $t$, for $\varphi=0.58$.}
  \label{fig:comparing-chi4-t-epsilon-t-HARD}
\end{figure}

\begin{figure}[ht!]
  \centering
  \includegraphics[width=\columnwidth]{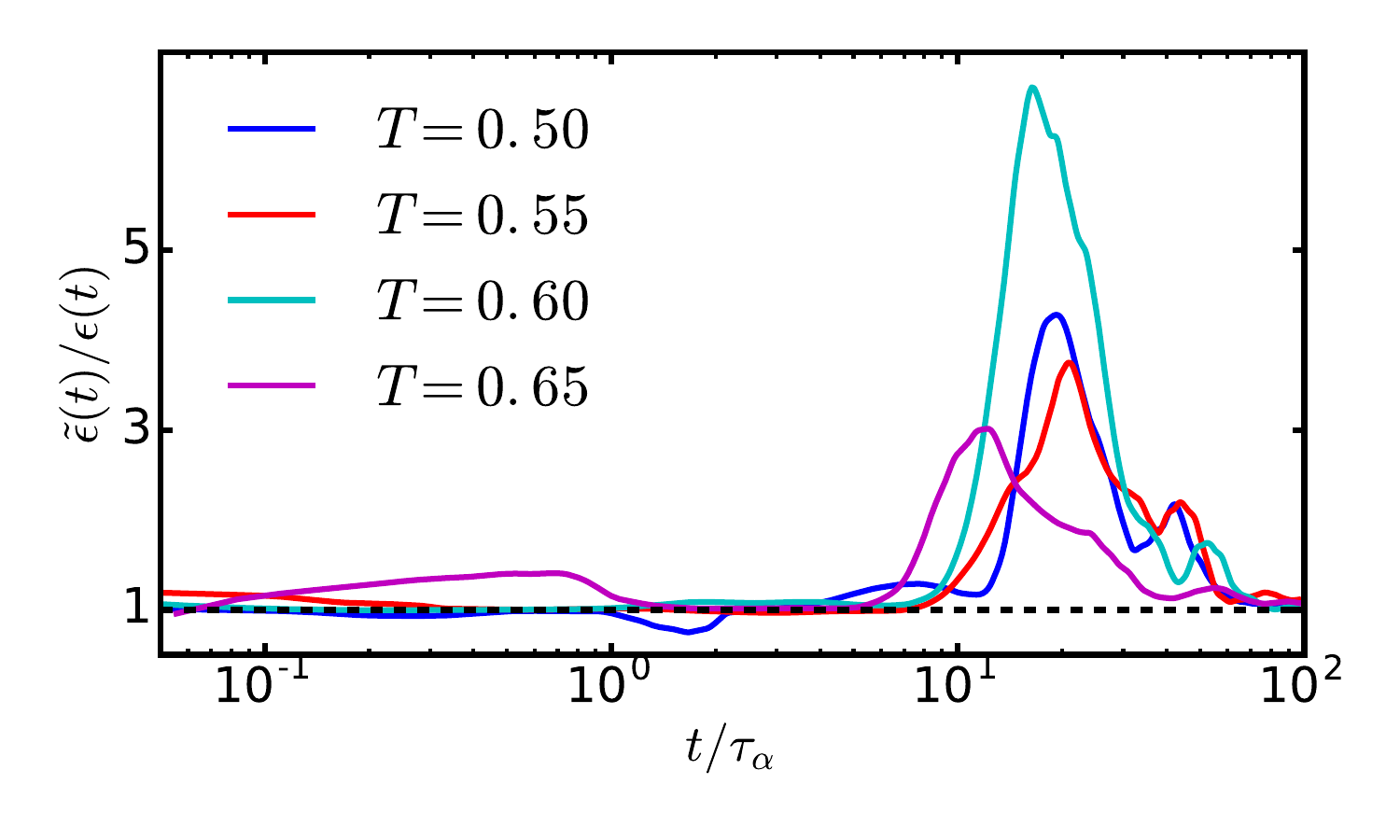}\\[-1.7ex]
  \includegraphics[width=\columnwidth]{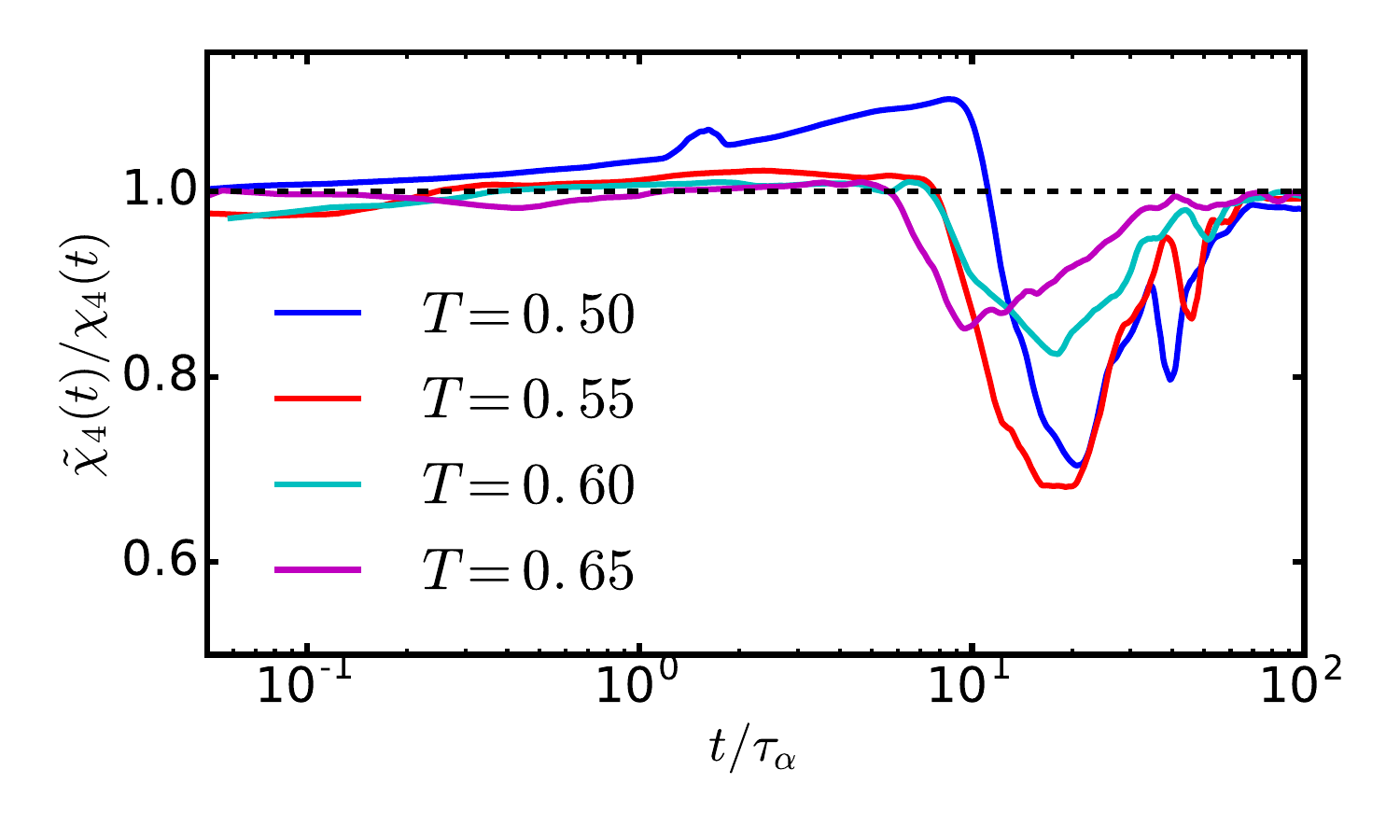}\\[-2.7ex]
 \caption{Effects of allowing for a $q$-independent background
   contribution $\chi_{4,b}(t)$ to $S_4(\vec{q},t)$ due to single
   particle fluctuations and initial density fluctuations, in the case
   of KALJ. $\chi_4(t)$ [$\tilde{\chi}_4(t)$] is the dynamic
   susceptibility and $\epsilon(t)$ [$\tilde{\epsilon}(t)$] is the
   rms fitting error per degree of freedom in the interval $0 < q <
   q_{\epsilon}=0.55$, obtained from a fit of $S_4(\vec{q},t)$ vs $q$
   allowing [not allowing] for a background contribution.  Top panel:
   $\tilde{\epsilon}(t)/\epsilon(t)$ vs time $t$, for
   $T=0.50,0.55,0.60,0.65$. Bottom panel:
   $\tilde{\chi}_4(t)/\chi_4(t)$ vs time $t$, for
   $T=0.50,0.55,0.60,0.65$.}
  \label{fig:comparing-chi4-t-epsilon-t-KALJ}
\end{figure}

The form for $S_4(\vec{q},t)$ is fitted for each time separately in a
two-step procedure.  In the first step, a wide fitting range is used: $0
< q < q_M$ with $q_M \sim q_0/2 \approx \pi/r_{NN}$. We choose $q_M =
3.0$ and $q_M = 3.5$ for HARD and KALJ respectively. In this first step,
the $q$-independent background $\chi_{4,b}$ is determined. In the second
step, a much narrower range $0 <q< q_m \ll q_M$ is used, and
$\chi_{4,b}$ is now kept as a fixed value as determined in the first
step.  The fitting ranges for the second fit are $0<q<q_m = 0.7$ and
$0<q<q_m = 0.8$ for HARD and KALJ respectively. The four independent
simulation runs are fitted separately for each value of the control
parameter. The average results and statistical errors of the fits are
calculated as the average and the standard deviation of the average from
those four fits. The LOESS smoothing technique (Ref.~\cite{Cleveland1979})
is used to reduce noise in the reported results for $\chi_4^{\rm
  cr}(t)$, $\xi_4^{\rm cr}(t)$, and $c(t)$.
The values of $\xi_4^{\rm cr}(t)$ for HARD determined with this
procedure are somewhat sensitive to the range of wavevectors used in
the second step of the fitting procedure. To quantify the size of this
effect, the second step discussed above is performed for $q_m \in \{
0.6, 0.7, 0.8, 0.9 \}$, and the systematic error bars due to the
choice of $q_m$, which are shown in Fig.~\ref{fig:xi_t}, are
evaluated for each time and packing fraction as the standard deviation
of the average of $\xi_4^{\rm cr}(t)$ over those four determinations.

\section{Effects of the presence of the $q$-independent background
  $\chi_{4,b}(t)$ on the determination of $\chi_4(t)$ and $\xi_4^{\rm cr}(t)$}
\label{sec:compare-fits}

\begin{figure}[ht]
  \centering
  \includegraphics[width=\columnwidth]{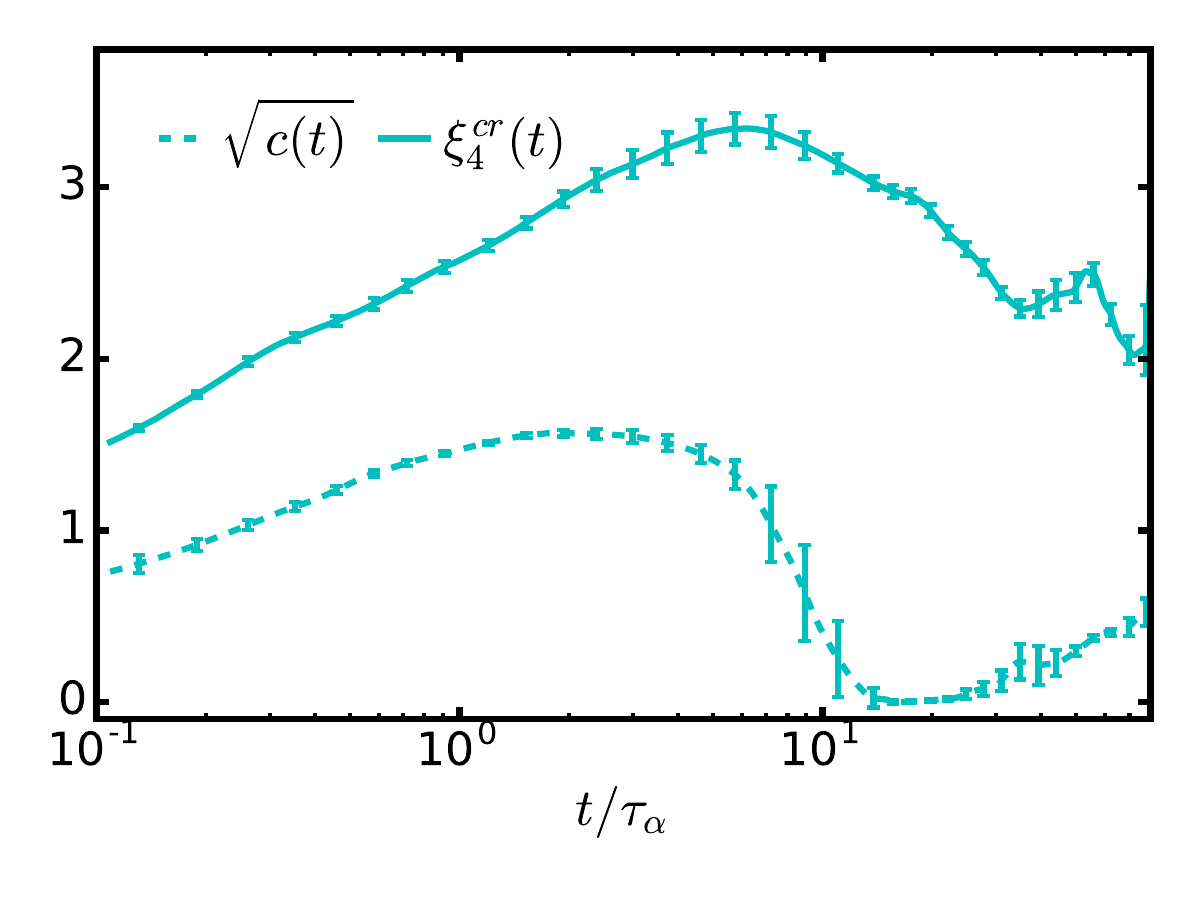}\\[-3.0ex]
  \caption{Time dependence of quartic term coefficient $\sqrt{c(t)}$ in
    the four-point function fit, for HARD at $\varphi=0.57$.
    $\xi_4^{\rm cr}(t)$ is shown for comparison, since both quantities
    have dimension of length.}\label{fig:sqrt-c_t} 
\end{figure}

The presence of the background term $\chi_{4,b}(t)$, due mostly to
single particle fluctuations and to initial density fluctuations, has a
strong effect on the determination of $S(\vec{q},t)$ for small
wavevector $q$, and consequently on the determination of $\chi_4(t)$ and
$\xi_4^{\rm cr}(t)$. Fig.~\ref{fig:comparing-fits-vs-q_s_4} shows an
example of those effects by comparing the determination of
$S_4(\vec{q},t)$ as a function of $q$ for HARD at $\varphi=0.58, t = 20
\tau_{\alpha}$ by using two different methods: one is a fit that allows
for a $q$-independent background $\chi_{4,b}(t) \ne 0$, consistent with
the decomposition introduced in this work; the other is a fit that
imposes the condition $\chi_{4,b}(t) = 0$. It is clear that outside a
narrow range of $q$ values where the two fits are equivalent, the one
that allows for a nonzero flat background is a much better
representation of the
data. Figs.~\ref{fig:comparing-chi4-t-epsilon-t-HARD}
and~\ref{fig:comparing-chi4-t-epsilon-t-KALJ} presents a more systematic
demonstration of the effects of the background term, for the hard-sphere
system and the Kob-Andersen Lennard-Jones system respectively. In these
figures, $\chi_4(t)$ [$\tilde{\chi}_4(t)$] is the dynamic
susceptibility, $\xi_4^{\rm cr}(t)$ [$\tilde{\xi}_4(t)$] is the dynamic
correlation length, and $\epsilon(t)$ [$\tilde{\epsilon}(t)$] is the rms
fitting error per degree of freedom in the interval $0 < q <
q_{\epsilon}$, obtained from a fit of $S_4(\vec{q},t)$ vs $q$ allowing
[not allowing] for a background contribution. In each figure, the first
panel from the top shows $\tilde{\epsilon}(t)/\epsilon(t)$ vs $t$, and
the second panel shows $\tilde{\chi}_4(t)/\chi_4(t)$ vs time $t$. In the
case of HARD, there is a third panel that shows
$\tilde{\xi}_4(t)/\xi_4^{\rm cr}(t)$ and $\hat{\xi}_4(t)/\xi_4^{\rm
  cr}(t)$ vs time $t$, for $\varphi=0.58$. Here $\hat{\xi}_4(t)$ is the
dynamic correlation length obtained from the following procedure: first,
$\chi_4(t)$ is obtained from a fit allowing for a nonzero background;
after that, $\chi_4(t)$ is kept fixed and a new fit is performed with
the background constrained to be zero, which produces the value of
$\hat{\xi}_4(t)$.  We notice that in all cases the rms fitting error is
either the same or smaller if the background term is allowed. In most
cases the difference becomes largest for times $t$ in the interval $10
\tau_{\alpha} < t < 100 \tau_{\alpha}$. For example, the ratio
$\tilde{\epsilon}(t)/\epsilon(t)$ is in the range of $2-8$ for HARD at
$\varphi=0.57,0.58$ at most times in that interval. For the same time
range, the effect on the determination of the dynamic susceptibility is
particularly large for HARD at $\varphi=0.58$, namely a reduction of up
to a factor of $\approx 2$ if the background is assumed to be zero. For
KALJ, the effect is strongest in the same time range, with a maximum
reduction by a factor of $\approx 1.4$ for $T=0.50,0.55$. For the
correlation length, there is a clear reduction in the value measured if
the background is ignored, which starts to be noticeable at $t \approx
10 \tau_{\alpha}$, and becomes gradually stronger as time
grows. Although slightly weaker for $\hat{\xi}_4(t)$ than for
$\tilde{\xi}_4(t)$, the effect is very similar in both cases, which
shows that it cannot be avoided just by constraining the fit by fixing a
better determined value of the dynamical susceptibility.

\section{Quartic term in the generalized Ornstein-Zernicke form.}
\label{sec:QuarticTerm}
The quartic coefficient $c(t)$ included in the denominator of the
generalized Ornstein-Zernicke fitting form in Eq.~\ref{eq:modified-oz}
allows the definition of a length $\sqrt{c(t)}$, which turns out to be
generally smaller than $\xi_4^{\rm cr}(t)$, as shown in
Fig.~\ref{fig:sqrt-c_t}.

\bibliographystyle{apsrev4-1}
\bibliography{collective-sp_pre} 

\end{document}